\documentclass[12pt]{iopart}
\usepackage{mathptmx}
\usepackage{graphicx}

\begin{document}

\title[Degenerate Ground State and Anomalous Flux Hysteresis in an YBCO SQUID]{Degenerate Ground State and Anomalous Flux Hysteresis in an YBa$_2$Cu$_3$O$_7$ Grain Boundary r.f. SQUID}

\author{C H Gardiner$^1$\footnote[1]{To whom correspondence should be addressed
(Before 1 May 2004: carol.gardiner@npl.co.uk; After 1 May 2004:
carol.webster@npl.co.uk).}, R A M Lee$^2$, J C Gallop$^1$, A Ya
Tzalenchuk$^1$, J C Macfarlane$^3$ and L Hao$^1$}

\address{$^1$National Physical Laboratory, Queens Road, Teddington, Middlesex, TW11 0LW, UK}

\address{$^2$California Institute of Technology, Condensed Matter Physics, Mail Code 114-36, 1251 E. California Blvd., Pasadena,  CA 91125, USA}

\address{$^3$Department of Physics, University of Strathclyde, Glasgow, G4 0NG, UK}

\begin{abstract}
We report measurements of the flux hysteresis curves and trapped
flux distribution in an YBa$_2$Cu$_3$O$_7$ r.f. SQUID containing
two closely spaced grain boundary Josephson junctions in parallel.
Broadening of the flux distribution from $T$ = 15\,K to 30\,K is
followed by a bifurcation at $T$ = 35\,K which corresponds to a
degenerate ground state.  Above $T \approx 40$\,K the bifurcation
disappears, the flux distribution narrows significantly and small
secondary loops appear in the hysteresis curves.  This behaviour
can be modelled qualitatively if we assume a temperature dependent
second harmonic term in the current-phase relationship of the
junctions.
\end{abstract}




\section{Introduction}

In recent years there has been much interest in Josephson
junctions and superconducting quantum interference devices
(SQUIDs) fabricated from high temperature superconductor (HTS)
materials.  A series of elegant experiments have utilised the
effects of phase continuity in superconducting rings
\cite{Wollman:1993, Tsuei:1994, Kirtley:1995} and quantum
interference in wide junctions \cite{Wollman:1995, Miller,
Brawner} to demonstrate the predominantly $d_{x^2 - y^2}$-wave
symmetry of the order parameter.

It has been shown that it is possible to trap half flux quanta in
rings containing an odd number of ``$\pi$-junctions"
\cite{Tsuei:1994}.  A $\pi$-junction is so called because in its
ground state the phase change of the order parameter across it is
$\pi$.  This occurs when the current-phase relationship (CPR) is
inverted (i.e. the critical current becomes negative) --- an
effect that occurs in junctions between two superconductors of
$d_{x^2 - y^2}$-wave symmetry when their crystal axes are
misaligned by more than 45$^{\circ}$ \cite{Sigrist}.  Such
junctions can be created by a number of techniques, including the
use of HTS grain boundaries.

It has also been shown that the CPR in HTS junctions is
non-sinusoidal, possessing a second harmonic component whose
magnitude depends on the relative orientation of the $d_{x^2 -
y^2}$-wave superconductors \cite{Zhang}, and whose sign can be
temperature dependent \cite{Tanaka}. For a junction contained in a
SQUID ring, the CPR can be expanded as a Fourier series in $\phi$,
the phase difference of the superconducting order parameter across
the junction:

\begin{equation}
I = I_{\rm c1}\sin(\phi) + I_{\rm c2}\sin(2\phi) + \ldots,
\label{eq:CPR}
\end{equation}

\noindent where $I$ is the supercurrent flowing through the
junction.  In an asymmetric 45$^{\circ}$ [001]-tilt grain boundary
junction between two $d_{x^2 - y^2}$-wave superconductors, the
first harmonic $I_{\rm c1}$ is expected to disappear by symmetry
\cite{Sigrist} (although in practice it does not disappear
completely, due to microscopic or nanoscale faceting of the
boundary \cite{Hilgenkamp}).  The second harmonic $I_{\rm c2}$
then becomes prominent, and has been observed in recent studies
\cite{Il'ichev:1998, Il'ichev:2001}.

In this paper we present measurements of the distribution of
trapped flux and of the flux hysteresis curves in an
YBa$_2$Cu$_3$O$_7$ SQUID containing asymmetric 30$^{\circ}$
[001]-tilt grain boundary junctions.  Our observations appear to
indicate the presence of a temperature dependent second harmonic
term.

\section{Device fabrication}

\begin{figure}[!ht]
\begin{center}
\includegraphics{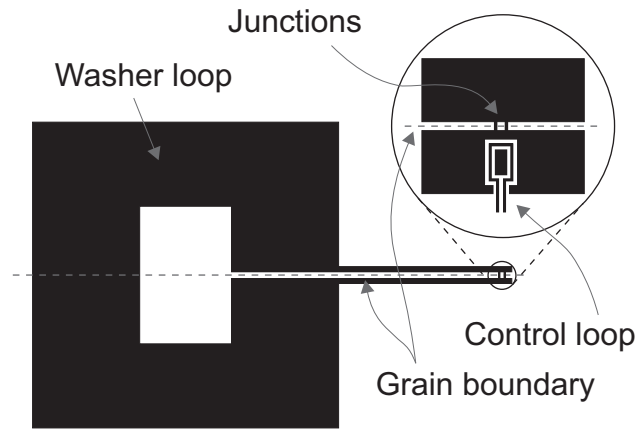}
\caption{Schematic diagram of washer and Josephson junctions (not
to scale).  The control loop is used to provide external magnetic
flux.}
\label{fig:ChipDiagram}
\end{center}
\end{figure}

The device consists of a thin film of YBa$_2$Cu$_3$O$_7$ ($\sim
200$\,nm thick), grown by pulsed laser deposition on a SrTiO$_3$
bi-crystal substrate with a well-defined asymmetric misorientation
angle of 30$^{\circ}$.  It was originally designed as a potential
dissipationless noise thermometer \cite{Gallop, Lee}, and
therefore consists of a large square washer (6\,mm $\times$ 8\,mm)
interrupted by two closely spaced grain boundary junctions in
parallel (5\,$\mu$m $\times$ 3\,$\mu$m).  A small control loop is
positioned close to the junctions to allow external flux to be
applied.  Figure \ref{fig:ChipDiagram} shows a schematic diagram
of the device. The incorporation of two parallel junctions is
historical, and is irrelevant to the work described in this paper.
Since the loop enclosed by the junctions is tiny in comparison to
the main washer loop, and hence has negligible self inductance, we
will treat the pair as a single junction.  The other Josephson
junction formed by the grain boundary on the wide part of the
washer loop has a very large critical current, and can therefore
be ignored. Hence, we have an r.f. SQUID consisting of a washer
effectively interrupted by a single grain boundary junction.

\section{Experimental details}

\begin{figure}[!ht]
\begin{center}
\includegraphics{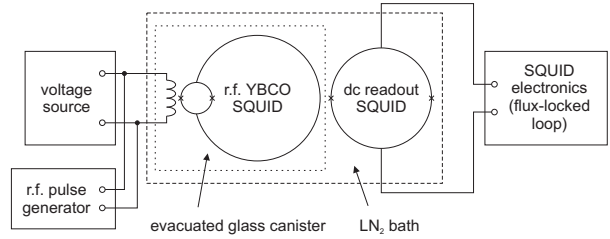}
\caption{Schematic diagram of the experimental setup.}
\label{fig:setup}
\end{center}
\end{figure}

A schematic diagram of the experimental setup is shown in Figure
\ref{fig:setup}.  Inside an evacuated glass canister the device
was mounted on the end of a sapphire rod in thermal contact with
the cold finger of a continuous flow He siphon. Temperature
control was achieved by passing high frequency a.c.
current\footnote{A low pass filter on the readout SQUID was used
to filter out any stray fields from the heater coil.} through a
wire wound heater. A separate d.c. readout SQUID was positioned
outside the glass, 6\,mm from the r.f. SQUID and well inductively
coupled. The whole system was immersed in a bath of liquid
nitrogen.

\section{Results}

\begin{figure}[!ht]
\begin{center}
\includegraphics{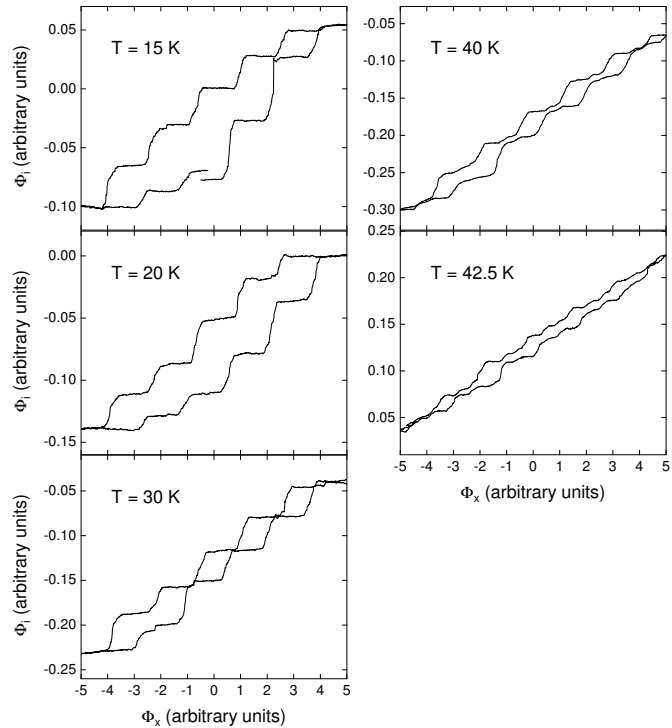}
\caption{Temperature dependence of hysteresis curves. Small
secondary loops appear at $T \ge 40$\,K.}
\label{fig:HysteresisMeasurements}
\end{center}
\end{figure}

The flux hysteresis curves were measured by slowly sweeping the
d.c. current through the control loop at each set temperature. The
internal flux threading the washer was monitored by the readout
SQUID, which was operated in flux-locked mode.  The measurements
are displayed in Figure \ref{fig:HysteresisMeasurements}.  We have
estimated the linear background arising from coupling of the
external applied field into the readout SQUID, and subtracted this
from each plot.

The data show discrete steps in the internal flux, which we
interpret as the admission of individual flux quanta into the
washer loop. However, the hysteresis curves do not correspond to
those of an ideal SQUID.  At temperatures of 15\,K, 20\,K and
30\,K the steps are not all exactly equal in height.  Some,
particularly those at the extreme ends of the field sweep, appear
to be approximately half the average size, suggesting that the
washer sometimes admits half flux quanta. At 40\,K small steps in
the internal flux begin to appear alongside the larger steps seen
at lower temperatures. The alternation of large and small steps
causes secondary loops to appear between the main loops of the
hysteresis curves.  The half-sized steps evident at $T=15$\,K to
$T = 20$\,K and the secondary loops at $T \ge 40$\,K are
suggestive of a second harmonic component in the CPR. At 42.5\,K
the secondary loops increase in size, suggesting an increase in
the importance of the second harmonic.

To sample the internal flux distribution we applied a series of
r.f. pulses to the control loop and used the readout SQUID to
monitor the internal flux threading the washer.  The principle
behind this procedure is as follows.  By carefully tuning the
amplitude and frequency of the r.f. signal it is possible to
suppress the critical currents of the junctions through the a.c.
Josephson effect.  Therefore, during a high r.f. pulse, the
potential barriers between the metastable flux states of the ring
are lowered, and the flux can fluctuate classically.  During a low
pulse, the barriers are restored and the flux is trapped in a
metastable quantised state.  Thus, a succession of high and low
r.f. pulses causes the internal flux threading the washer loop to
be repeatedly trapped and released.  By recording the signal from
the readout SQUID during each low r.f. pulse a histogram of flux
states can be built up.

To achieve the r.f. amplitude and frequency required to suppress
the junction critical currents we applied a continuous r.f. signal
while measuring the flux hysteresis curves.  When discrete flux
jumps no longer occurred we were able to deduce that the critical
currents were suppressed.  We then applied a series of r.f. pulses
at each set temperature and recorded the internal flux state of
the washer using the readout SQUID, the response of which we
binned into histograms.

\begin{figure}[!ht]
\begin{center}
\includegraphics{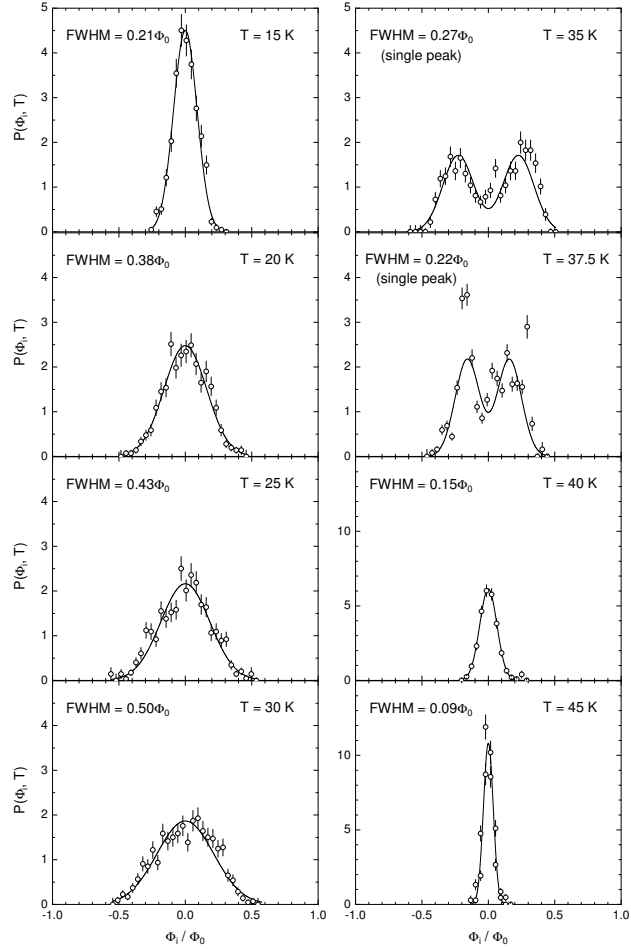}
\caption{Temperature dependence of flux distribution.  The open
circles are the normalised histogram data and the lines are
gaussian fits using one or two gaussians.  The distribution
bifurcates at $T = 35$\,K.}
\label{fig:FluxHistograms}
\end{center}
\end{figure}

From the low temperature hysteresis curves we were able to
determine the change in the readout SQUID signal brought about by
admission of a single flux quantum to the washer loop.  This
allowed us to put the histograms on an absolute flux scale. We
then normalised the data to an area of 1 for comparison with
theory. Figure \ref{fig:FluxHistograms} shows the measured flux
distibution at a series of temperatures.

The flux distribution is observed to broaden considerably from a
width of 0.21$\Phi_0$ at 15\,K to 0.5$\Phi_0$ at 30\,K.  It then
suddenly bifurcates at $T = 35$\,K.  The peaks of the bifurcated
distribution are separated by 0.5$\Phi_0$ at 35\,K, which clearly
indicates the presence of a strong second harmonic in the CPR.  At
37.5\,K the peaks are separated by 0.3$\Phi_0$, which suggests a
decline in the contribution of the second harmonic. At $T = 40$\,K
the flux distribution suddenly reverts to a single peak and
becomes much narrower than at lower temperatures.

\section{Modelling}

In this section we model the flux hysteresis curves and the flux
distribution, assuming a single Josephson junction and taking into
account the possibility of a second harmonic term in the CPR.

The internal flux threading a superconducting ring is given by

\begin{equation}
\Phi_{\rm i} = \Phi_{\rm x} + IL,
\label{eq:PhiI}
\end{equation}

\noindent where $\Phi_{\rm x}$ is the external applied flux, $I$
is the supercurrent and $L$ is the self-inductance of the ring.
The supercurrent can be obtained from the CPR (\ref{eq:CPR}). The
requirement of phase continuity around a superconducting ring
gives rise to the condition $\phi + \frac{2\pi\Phi_{\rm
i}}{\Phi_0} = 2n\pi$ (quantisation of the fluxoid
\cite{GallopBook}).  If we substitute this into (\ref{eq:CPR}) we
obtain

\begin{equation}
I = -I_{\rm c1}\sin\left(\frac{2\pi\Phi_{\rm i}}{\Phi_0}\right)
-I_{\rm c2}\sin\left(\frac{4\pi\Phi_{\rm i}}{\Phi_0}\right) -
\ldots. \label{eq:I}
\end{equation}

\noindent Substituting (\ref{eq:I}) into (\ref{eq:PhiI}) and
rearranging, we obtain

\begin{equation}
\Phi_{\rm x} = \Phi_{\rm i} + I_{\rm
c1}L\sin\left(\frac{2\pi\Phi_{\rm i}}{\Phi_0}\right) + I_{\rm
c2}L\sin\left(\frac{4\pi\Phi_{\rm i}}{\Phi_0}\right).
\label{eq:PhiI-vs-PhiX}
\end{equation}

\noindent Equation (\ref{eq:PhiI-vs-PhiX}) allows us to model the
hysteresis curves by varying the magnitudes and signs of the
parameters $I_{\rm c1}$ and $I_{\rm c2}$.

We model the flux distribution according to the Boltzmann
distribution, so we must first calculate the internal energy of
the ring. This is given by

\begin{equation}
U(\Phi_{\rm i}, \Phi_{\rm x}) = \frac{(\Phi_{\rm i} - \Phi_{\rm
x})^2}{2L} + \int_0^{\phi} I\,\rm d\phi,
\end{equation}

\noindent where the first term is the classical free energy of the
ring and the second term is the Josephson coupling energy of the
junction.  Evaluating the integral and again utilising the
condition $\phi + \frac{2\pi\Phi_{\rm i}}{\Phi_0} = 2n\pi$ we
obtain

\begin{equation}
U(\Phi_{\rm i}, \Phi_{\rm x}) = \frac{(\Phi_{\rm i} - \Phi_{\rm
x})^2}{2L} - \frac{I_{\rm
c1}\Phi_0}{2\pi}\cos\left(\frac{2\pi\Phi_{\rm i}}{\Phi_0}\right) -
\frac{I_{\rm c2}\Phi_0}{4\pi}\cos\left(\frac{4\pi\Phi_{\rm
i}}{\Phi_0}\right).
\end{equation}

\noindent  This function has metastable minima if $I_{\rm c1}$ and
$I_{\rm c2}$ are large enough.  The flux distribution is given by

\begin{equation}
P(\Phi_{\rm i}, \Phi_{\rm x}, T) = P_0\exp\left(-\frac{U(\Phi_{\rm
i}, \Phi_{\rm x})}{k_{\rm B}T}\right),
\label{eq:P}
\end{equation}

\noindent where $P(\Phi_{\rm i}, \Phi_{\rm x}, T)$ is the
normalised probability of finding the ring in flux state
$\Phi_{\rm i}$, at temperature $T$, given an external flux
$\Phi_{\rm x}$.  $P_0$ is the constant of proportionality
(obtained from the condition $\int_{-\infty}^{\infty} P(\Phi_{\rm
i}, \Phi_{\rm x}, T)\,\rm d\Phi_{\rm i} = 1$) and $k_{\rm B}$ is
Boltzmann's constant.

By varying the parameters $I_{\rm c1}$, $I_{\rm c2}$ and $L$ in
Equations (\ref{eq:PhiI-vs-PhiX}) and (\ref{eq:P}) we can model
the observed flux hysteresis curves and the flux distributions.  A
wide variety of different behaviour can be reproduced by
considering both negative and positive values of $I_{\rm c1}$ and
$I_{\rm c2}$:

\begin{itemize}
\item Half-sized steps or secondary loops in the hysteresis curves
require $I_{\rm c2}$ to be negative and $I_{\rm c1}$ to be either
negative or positive.  If both are large, we obtain large
hysteresis loops with half-sized steps at the extreme ends.  If
both are smaller we obtain small hysteresis loops interspersed
with even smaller secondary loops.

\item A bifurcated flux distribution with peak separation $\le
0.5\Phi_0$ can be achieved most readily with negative $I_{\rm
c2}$.  Negative $I_{\rm c1}$ gives rise to a bifurcation of the
flux distribution with peak separation $\le \Phi_0$.

\item A single peak in the flux distribution can be achieved with
a wide variety of different parameters.  Positive values of
$I_{\rm c1}$ and/or $I_{\rm c2}$ result in a narrow peak, whereas
small negative values of $I_{\rm c1}$ and/or $I_{\rm c2}$ result
in a broad peak.  Increasing $L$ increases the peak width for all
values of $I_{\rm c1}$ and $I_{\rm c2}$.
\end{itemize}

\begin{figure}[!ht]
\begin{center}
\includegraphics{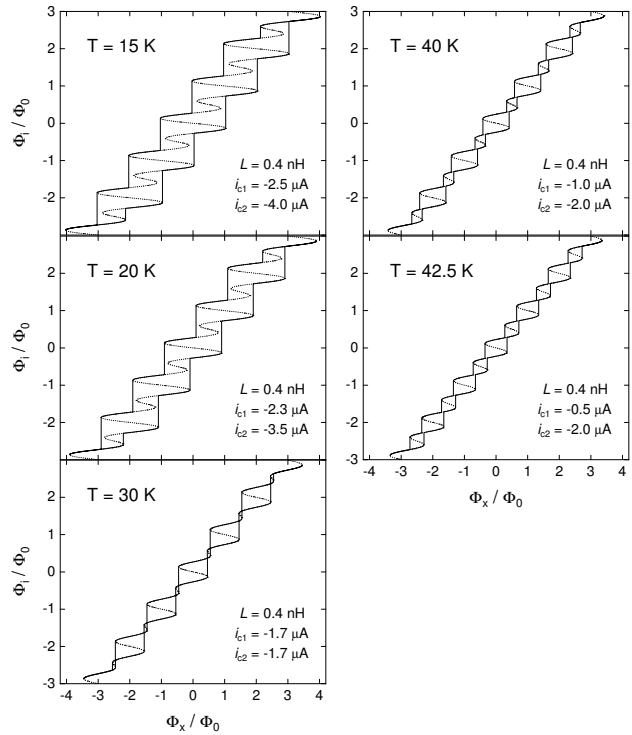}
\caption{Simulated hysteresis curves (solid lines). The dashed
lines show regions of the $\Phi_{\rm i}$ versus $\Phi_{\rm x}$
curve that are inaccessible due to the potential barrier between
metastable flux states.}
\label{fig:HysteresisSimulations}
\end{center}
\end{figure}

\begin{figure}[!ht]
\begin{center}
\includegraphics{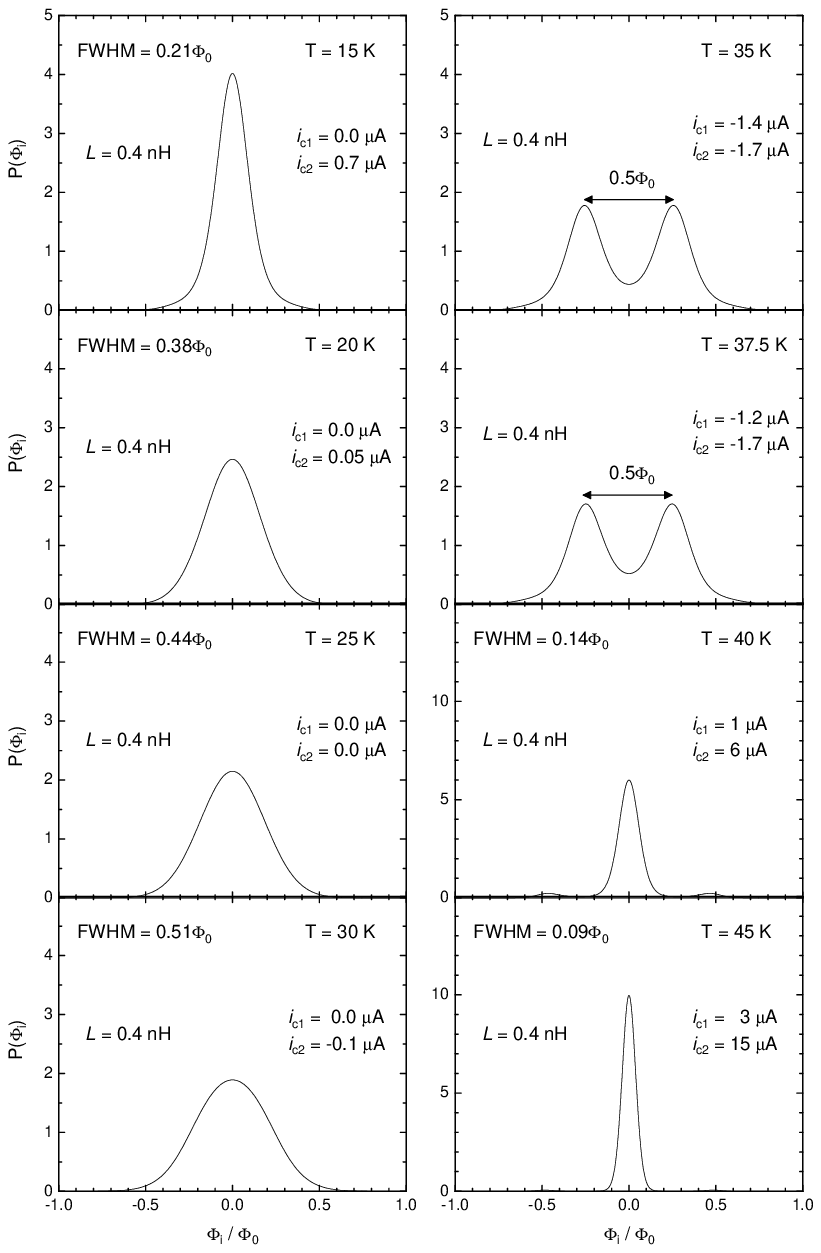}
\caption{Simulations of the flux distribution.  The bifurcation at
35\,K can be reproduced with a large negative $I_{\rm c2}$.  The
very narrow single peak at $T \ge 40$\,K can be reproduced with a
large positive $I_{\rm c2}$.}
\label{fig:FluxSimulations}
\end{center}
\end{figure}

Figure \ref{fig:HysteresisSimulations} shows simulations of the
hysteresis curves which best model the data shown in Figure
\ref{fig:HysteresisMeasurements}. Figure \ref{fig:FluxSimulations}
shows simulations of the flux distributions which best model the
data shown in Figure \ref{fig:FluxHistograms}.  A value of $L =
0.4$\,nH was used in the simulations of both the hysteresis curves
and the flux distributions, as this ensured insignificant
population of higher flux states, in accordance with our
observations.  We also assumed no external flux to be present.

Qualitative agreement is obtained between measurement and
simulation for both the hysteresis curves and the flux
distribution.  The half-sized steps at the extreme ends of the
flux hysteresis curves at $T$ = 15\,K and $T$ = 20\,K can be
reproduced using negative values for both $I_{\rm c1}$ and $I_{\rm
c2}$. The magnitudes of $I_{\rm c1}$ and $I_{\rm c2}$ decrease
with temperature, causing the large hysteresis loops to gradually
close up and the secondary loops to appear. The broadening of the
flux distribution can be modelled with a value of $I_{\rm c2}$
which starts small and positive, but decreases through zero to
become negative for $T \ge 30$\,K.  At $T = 35$\,K and $T =
37.5$\,K the bifurcation in the flux distribution is reproduced
with much larger negative values of $I_{\rm c2}$. For $T \ge
40$\,K the flux distribution recombines to form a single, narrow
peak.  This can be reproduced using large positive values of
$I_{\rm c2}$.

Although we have obtained qualitative agreement between the model
and the two data sets, we have not been able to find sets of
parameters that are consistent with both data sets at all
temperatures.  Over the temperature range 30--42.5\,K we model the
hysteresis curves using negative values of $I_{\rm c1}$ and
$I_{\rm c2}$, where $I_{\rm c2}$ increases slightly in magnitude,
but $I_{\rm c1}$ decreases.  Over the temperature range
35--37.5\,K we are able to model the bifurcated flux distributions
using very similar parameters. However, outside this temperature
range the parameters required to model the single peaks observed
in the flux distributions are very different from those used for
the hysteresis curves.

Unfortunately no hysteresis curves were measured at $T$ = 35\,K
and $T$ = 37.5\,K, so the results of the two simulations cannot be
compared directly.  At $T$ = 40\,K we do have data for both the
hysteresis curve and the flux distribution, but the parameters
required to model the two sets of data are quite different:
negative $I_{\rm c1}$ and $I_{\rm c2}$ for the hysteresis curve,
but positive $I_{\rm c1}$ and $I_{\rm c2}$ for the flux
distribution.

Over the temperature range $T$ = 15--30\,K we also have data for
both hysteresis curves and flux distributions, but again the
parameters used to model these are very different.  We also note
that our simulations of the hysteresis curves at $T$ = 15\,K and
$T$ = 20\,K contain six flux jumps rather than the five observed.
If we had used positive values of $I_{\rm c1}$ rather than
negative ones, we would have obtained simulations of almost
identical shape, but with five steps.

\begin{figure}[!ht]
\begin{center}
\includegraphics{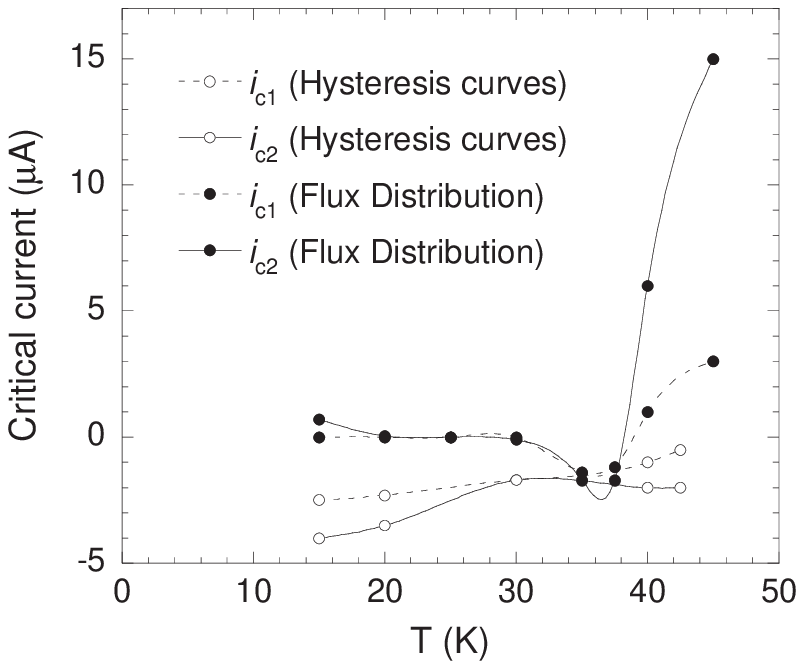}
\caption{Temperature dependence of the values of $I_{\rm c1}$ and
$I_{\rm c2}$ obtained from simulations of the hysteresis curves
and the flux distributions.  The lines are a guide to the eye.}
\label{fig:CriticalCurrents}
\end{center}
\end{figure}

Figure \ref{fig:CriticalCurrents} shows the temperature dependence
of the parameters $I_{\rm c1}$ and $I_{\rm c2}$ obtained from the
two simulations.  The only temperature range over which the two
parameter sets agree is 35--37.5\,K.

\section{Discussion}

Before concluding, we will discuss other possible mechanisms which
could explain our data.  A doubly degenerate ground state, as
observed from the bifurcation of the flux distribution, can be
obtained by a number of mechanisms.  A superconducting ring that
contains an odd number of $\pi$-junctions is expected to produce a
degenerate ground state regardless of the existence of a second
harmonic term in the CPR \cite{Copetti}. However, the two
degenerate flux states are expected to be separated by $\sim
\Phi_0$ in such a ring, whereas we observe a separation of $\sim
0.5\Phi_0$.  In order to achieve a separation of $\le 0.5\Phi_0$
using only a negative $I_{\rm c1}$ we find that the critical
current must be very low, leading to a very small potential
barrier between the degenerate states, and therefore, poorly
resolved flux states.  We observe well separated flux states,
which can only be modelled using a substantial negative $I_{\rm
c2}$.

If we had overestimated by a factor of 2 the response of the
readout SQUID to a flux change of $\Phi_0$ in the washer loop,
this would mean that the peaks in the bifurcated flux distribution
were separated by a factor of $\Phi_0$ rather than $0.5\Phi_0$.
However, we feel that this is unlikely since it would mean that
the majority of the flux jumps observed in the hysteresis curves
were $2\Phi_0$.

The ground state of a superconducting ring is also doubly
degenerate in the presence of an external flux of $\Phi_0/2$.
However, in such a case the degenerate states are separated by
exactly $\Phi_0$, which is not observed.  We also think it highly
unlikely that any stray external flux should take a value of
exactly $\Phi_0/2$. If the stray flux should deviate from this
value, we would expect one state to be preferred, but to within
experimental uncertainty we observe the two states to be equally
populated.

Finally, it is possible that the secondary loops observed in the
hysteresis curves are due to the geometry of the r.f. SQUID. It is
unfortunate that the historical design incorporated two junctions
in parallel rather than one, and we cannot rule out the
possibility that flux becomes trapped in the small loop enclosed
by the two junctions.  However, the difference in area, and
therefore in self inductance, between this small loop and the main
washer loop is so large that we feel this is unlikely.  It would
be worthwhile to repeat the measurements using a ring containing
only one junction.

\section{Conclusion}

We have measured the flux hysteresis curves and the distribution
of trapped flux in a thin film YBa$_2$Cu$_3$O$_7$ SQUID consisting
of a large washer, interrupted by an effectively single [001]-tilt
grain boundary Josephson junction with an asymmetric
misorientation angle of 30$^{\circ}$.  Our measurements show
striking anomalies, namely the appearance of secondary loops in
the hysteresis curves and a bifurcation of the flux distribution.
Both anomalies are temperature dependent.

We have modelled our results using a simple model of the flux
dynamics, adapted to incorporate the effect of a possible second
harmonic in the CPR.  By careful choice of the parameters $I_{\rm
c1}$, $I_{\rm c2}$ and $L$ we obtain good qualitative agreement
between the data and simulations.  The bifurcation in the flux
distribution and the secondary loops in the hysteresis curves are
both modelled well with a large negative $I_{\rm c2}$ and a
smaller negative $I_{\rm c1}$.  However, at other temperatures we
have been unable to find a consistent set of parameters that
describe both the hysteresis curves and the flux distributions.

We conclude that we have observed tentative evidence for a
temperature dependent second harmonic term in the CPR.  Our
results are surprising because substantial second harmonic
components in the CPR have only been observed previously in grain
boundary junctions with misorientation angles of 45$^{\circ}$
\cite{Il'ichev:1998, Il'ichev:2001}.  It is expected that in such
junctions the first harmonic should disappear by symmetry.
However, in a 30$^{\circ}$ grain boundary junction, the first
harmonic is not expected to disappear, making our observation of
the second harmonic quite striking.  Previous measurements made by
us on a grain boundary SQUID with misorientation angle
24$^{\circ}$ revealed no evidence of a second harmonic term.
However, the critical currents of the junctions in that device
were so large that it was extremely difficult to suppress them
enough to measure the distribution of trapped flux.  Measurements
by Il'ichev et al.\ \cite{Il'ichev:1999} on a grain boundary SQUID
with misorientation angle 36$^{\circ}$ also revealed no evidence
of a second harmonic term.  However, we believe that our
conclusion does not contradict this, as the relative strengths of
the first and second harmonic terms are likely to be governed by
many microscopic factors such as faceting, homogeneity and
transparency of the grain boundary, which make every device
unique.

\ack{We would like to acknowledge the work of Derek Peden and the
University of Strathclyde where the devices were designed and
fabricated.  We also acknowledge the ESF pi-shift programme for
assisting one of us (CHG) in attending EUCAS 2003.  The
experimental work was funded by the DTI Quantum Programme.}

\end{document}